\documentstyle[11pt,aaspp4,flushrt]{article}

\slugcomment{accepted September 2, 1997}

\def\OIII{[{\ion{O}{3}}]} 
 
\def\Ha{{H$\alpha$}\/} \def\Hb{{H$\beta$}\/}
\def\C977{{\ion{C}{3}}~$\lambda$977\/} \def\Ciii{{\ion{C}{3}}]
$\lambda$1909\/} \def\Cii{{\ion{C}{2}}] $\lambda$2326\/}
\def\Civ{{\ion{C}{4}}~$\lambda$1549\/}
\def\Sii{[{\ion{S}{2}}]~$\lambda\lambda$6717,31\/}
\def\Siii{[{\ion{S}{3}}]~$\lambda\lambda$9069,9532\/}
\def\N991{{\ion{N}{3}}~$\lambda$991\/}
\def\Niii{{\ion{N}{3}}]~$\lambda$1750\/}
\def\Nii{[{\ion{N}{2}}]~$\lambda$6583\/}

\def\Heii{{\ion{He}{2}}~$\lambda$1640\/}
\def\He4686{{\ion{He}{2}}~$\lambda$4686\/}
\def\Oiii{{[{\ion{O}{3}}]~$\lambda$5007}\/}
\def\Oii{[{\ion{O}{2}}]~$\lambda$3727\/}
\def\Oi{[{\ion{O}{1}}]~$\lambda$6300\/}
\def\O4363{[{\ion{O}{3}}]~$\lambda$4363\/}
\def\Nev{[{\ion{Ne}{5}}]~$\lambda$3426\/}
\def\Neiii{[{\ion{Ne}{3}}]~$\lambda$3869\/}
 \def\kms{{km~s$^{-1}$}\/}

\def\ergs.cm2.s{{ergs~cm$^{-2}$~s$^{-1}$}\/}

\lefthead{Allen et al.}  \righthead{UV diagnostics for the emission
line gas in active galaxies}

\begin{document}

\title{UV DIAGNOSTICS FOR THE EMISSION LINE GAS IN ACTIVE GALAXIES }

\author{\sc Mark G. Allen\altaffilmark{1}, 
            Michael A. Dopita\altaffilmark{1}, and 
            Zlatan I. Tsvetanov\altaffilmark{2} }

\altaffiltext{1}{Mount Stromlo and Siding Spring Observatories, The
        Australian National University, Private Bag Weston Creek P.O.,
        ACT 2611, Australia; mga, mad@mso.mso.anu.edu.au}

\altaffiltext{2}{Department of Physics and Astronomy, Johns Hopkins
        University, Baltimore MD 21218, USA; zlatan@pha.jhu.edu}

\begin{abstract}

  Optical diagnostic diagrams are frequently ambiguous as a test of
  the photoionization or fast shock models of the narrow line regions
  of active galaxies. Here, we present a set of UV line ratio diagrams
  which can discriminate between pure shock and photoionization modes
  of excitation, and to some extent, also discriminate shocks with
  ionized precursors from photoionization.  These diagrams use
  relatively bright emission lines and reddening insensitive ratios
  and provide a practical observational test for separating the
  excitation mechanisms of the narrow line regions of active galaxies.
  The most useful diagrams are those involving the various ionization
  stages of Carbon, (\Oiii/\Hb) $vs.$ (\Civ/\Heii) and the purely UV
  ratio pair (\Cii/\Ciii) $vs.$ (\Civ/\Ciii). Temperature sensitive
  FUV lines \C977\ and \N991\ also provide good discriminants. The
  models are compared to observations of nearby AGN, and also to high
  redshift objects where the UV lines are shifted into the optical.

\end{abstract}

\keywords{galaxies:nuclei---galaxies:Seyfert---ultraviolet:galaxies}

\section{INTRODUCTION}

  Our knowledge of the energetics of Active Galactic Nuclei (AGN) is
  largely derived from observational interpretation of the effects of
  the nucleus on its environment. In particular, studies of the
  mechanisms responsible for the ionization of the ``narrow line
  region'' (NLR) are important for understanding both the way the
  nucleus releases its energy, and in deriving the physical parameters
  of the active nucleus itself.
  
  Energetic processes occurring close to the central black hole of an
  active galaxy are thought to give rise to a powerful ionizing
  continuum.  These processes may include emission from a hot
  accretion disk and/or synchrotron emission, and subsequent
  Comptonization of highly energetic electrons, all of which may
  result, to first approximation, in a power law spectrum of the UV
  and X-ray ionizing photons. A power-law spectrum of photons is also
  suggested by the shape of the UV continuum observed in many
  moderately high redshift quasars and by the extrapolation of the UV
  continuum to the X-ray region of the spectrum. In photoionization
  models of the NLR, the gas is considered to be illuminated by the
  ionizing radiation from the nucleus producing the observed emission
  line spectrum.  This scenario appeals to the unified schemes for AGN
  in which the nuclear engine and the broad line region are surrounded
  by an optically thick torus (\cite{ant85} 1985).  Well defined
  bi-conical shaped extended narrow line regions (ENLR) observed in
  some Seyfert 2s ( \cite{pog89} 1989; \cite{tt89} 1989; \cite{wil94}
  1994; \cite{sim97} 1997) have been quoted as supporting evidence for
  this picture, the bi-conical shape being the result of shadowing of
  ionizing radiation from the nucleus by an optically thick torus.
  
  Shocks provide an alternative mechanism for the ionization of the
  NLR based on the input of mechanical energy. High velocity shocks
  generate a powerful local UV radiation field which can ionize the
  gas and emit a highly excited emission line spectrum like that
  observed in the NLR.  There is little doubt that shocks exist in the
  NLR and are important (at least) in determining the phase structure
  of the gas and in generating the outflows observed in a number of
  Seyfert galaxies. For example, \cite{col96} (1996) find that 6 out
  of a sample of 22 edge on Seyfert galaxies show morphological
  and/or kinematic evidence for large scale outflows along the minor
  axis.  Turbulence and instabilities in these flows must generate
  shocks. These may arise as either bow shocks at the working surface
  of a jet or outflow, cloud shocks around inhomogeneities caught up
  in the outflow region, or as wall shocks generated by the
  development of Kelvin-Helmholtz instabilities (\cite{sut93} 1993) or
  else produced within hypersonic shearing entrainment regions
  (\cite{dop97} 1997).

  The relative importance of shocks compared with photoionization in
  determining the overall energetics of the NLR has not yet been
  established. An hypothesis advanced by \cite{dop95} (1995, 1996;
  hereafter DS95 and DS96 respectively) is that emission in the NLR
  may be entirely due to shocks.  Further evidence to support this
  concept has been adduced by \cite{bic97} (1997) in the case of
  powerful Gigahertz-peak spectrum (GPS), compact steep spectrum
  (CSS), and compact symmetric objects (CSO).  Seyfert galaxies are
  however radio quiet, and equipartition arguments suggest that the
  pressure in the relativistic plasma is inadequate to drive
  high-velocity shocks in the interstellar medium (\cite{wil88} 1988).
  Nonetheless there is a clear correlation of optical and radio
  morphologies in Seyfert galaxies (\cite{wil94} 1994; \cite{cap96}
  1995, 1996).  The apparent shortfall of energy (as traced by the
  radio synchrotron emission) in the radio lobes available to drive
  the optical emission via shocks, may be altered if mass entrainment
  into the relativistic jet maintains a large fraction of the energy
  of the jet in a thermal phase undetected at radio wavelengths as
  suggested by \cite{bic97} (1997).
  
  Traditionally, attempts to distinguish between different excitation
  mechanisms have relied on observations of emission lines at optical
  wavelengths. For example \ion{H}{2} regions, planetary nebulae and
  active galaxies can be distinguished from one another on the optical
  line ratio diagrams \Nii/\Ha\ $vs.$ \Oiii/\Hb\ and of \Oii/\Oiii\
  $vs.$ \Oiii/\Hb, \Nii/\Ha\ or \Oi/\Ha\ as described by
  \cite{bal81} (1981, hereafter BPT).  This scheme was revised by
  \cite{vei87} (1987) avoiding the use of the reddening sensitive
  \Oii/\Oiii\ ratio and setting down functional criteria for the
  choice of line ratios for diagnostic purposes. In particular, ratios
  should be made up of strong lines that are easy to measure in
  typical spectra, blended lines should be avoided, and the wavelength
  separation between the two lines should be small as possible so the
  ratio is relatively insensitive to reddening correction and flux
  calibration.  In addition one should avoid ratios of forbidden lines
  of different elements which lead to problems of abundance ratio
  dependence.

  The physical basis for this classification scheme is that the size
  of the partially ionized zone in which the low ionization lines
  \Nii, \Sii\ and \Oi\ preferentially arise, depends strongly on the
  nature of the object.  This zone is quite extended in objects
  photoionized by a spectrum containing a large fraction of high
  energy photons (such as a power-law), but is very thin when
  photoionized by OB stars which emit few photons with energy above 4
  Ryd.

  These optical tests have been unable to provide conclusive
  discrimination between shocks and photoionization because both of these 
  mechanisms can reproduce most of the observed optical line
  ratios (DS96 and references therein). However, the intensities of
  the UV lines are predicted to be much stronger in shocks than in
  simple photoionization models. This is because the UV collisionally
  excited lines such as \Civ, \C977, \N991\ emit strongly at the high
  temperatures ($2\times10^4-10^5$ K) in the cooling zone behind high
  velocity (150$-$500 \kms) shocks compared to photoionized plasma in
  which these species are excited at a temperature of $\sim10^4$ K.
  The large differences in the rates of collisional excitation of
  these UV lines therefore provides a potential means to discriminate
  between the models.

  The high electron temperatures generated in shocks compared to
  photoionization, can also be used to provide discriminants between
  the models.  Electron temperatures indicated by the \O4363/ \Oiii\
  ratio typically exceed the predictions of single zone
  photoionization models by several thousand degrees (\cite{tad89}
  1989). The common interpretation of this is that an extra local
  source of heating such as shocks is required. Indeed, DS95 have
  shown that shocks come closer to reproducing the observations of
  this ratio in AGN.  However, \O4363\ is a weak line and flux
  measurements in the literature may be inaccurate due to poor
  subtraction of the stellar continuum.\footnote{\cite{sto96} (1996)
  have recently described an accurate procedure for stellar
  subtraction in the region of the \O4363\ and \He4686 lines.} Also
  recent multi-zone photoionization models are able to produce higher
  temperatures so this diagnostic is rendered somewhat more ambiguous.
  DS95 suggest that the definitive test between the models may lie
  with more sensitive, temperature dependent line ratios such as
  \Ciii/\C977\ and \Niii/\N991.
  
  In this paper we demonstrate with the aid of line ratio diagrams,
  how the UV lines ( \N991, \Niii, \C977, \Civ, \Ciii, \Cii, \Heii)
  alone, and in combination with the optical line ratios \Oiii/\Hb\
  and \Nev/\Neiii\ can be used to discriminate between current shock
  and photoionization models and therefore provide a useful
  observational test of the excitation mechanism.  The models used are
  the shock models of DS96, a set of single zone photoionization
  models (ionization parameter $U$ sequences) calculated with the same
  code as the shock models, and the recent $A_{M/I}$ sequence from
  \cite{bin96} (1996, hereafter referred to as BWS96).  We have chosen
  reddening insensitive ratios of relatively bright lines that are
  accessible with instruments such as the $HST$ Faint Object
  Spectrograph (FOS) and the Space Telescope Imaging Spectrograph
  (STIS), so that these diagrams can be used as a practical
  observational tool to probe the physics of the NLR.
  
  This paper is organized as follows: The various models used for
  comparison are described in section 2.  Section 3 describes the
  utility of UV lines as diagnostics and the rationale for the
  selection of line ratios, followed by plots of the model grids on
  line ratio diagrams. We discuss the diagnostic capabilities of each
  of the diagrams, and also compare the model grids to a small set of
  observations, where both optical and UV line intensities have been
  measured. Also we consider diagnostics that are useful for
  high-$z$ AGN where the UV lines have been shifted into the optical
  band.

\section{THE MODELS}

\subsection{Photoionization}

  The simplest photoionization models one can construct for the NLR
  involve a planar slab of gas which is illuminated by a power-law
  ($F_\nu\propto\nu^{\alpha}$) spectrum of ionizing radiation. The
  intensity of the radiation is specified by the ionization parameter,
  $U$, defined as the ratio of the density of impinging photons to the
  outer number density of the cloud\footnote{$U=\frac{1}{c
  n_H}\int_{\nu_0}^{\infty}\frac{F_{\nu}\ d\nu}{h\nu}$ where $F_{\nu}$
  is the incident monochromatic ionizing flux, $\nu_0$ the ionization
  potential of H, $n_H$ the density at the front face of the
  illuminated gas, and $c$ the speed of light.}.  Typically sequences
  of models in which $U$ varies over the range $10^{-4}-10^{-1}$ are
  compared to NLR spectra. These models have been remarkably
  successful at reproducing most of the observed optical line ratios
  (\cite{fer86} 1986, \cite{vei87} 1987, \cite{rob87} (1987),
  \cite{bin88} 1988).  The difficulties with single zone models have
  recently been reviewed by BWS96. Problems include a too low \OIII\
  temperature and the weak UV lines compared to the observations. In
  order to produce strong UV lines, high electron temperatures and
  strong coronal lines, the emissivity of the optically thin zone
  which gives rise to the higher ionization lines, must be raised
  relative to the optically thick zone.  This can be done in composite
  models which include both optically thick (ionization bounded)
  clouds and optically thin (matter bounded) clouds, by increasing the
  solid angle subtended by the optically thin clouds relative to the
  optically thick clouds. In this fashion, BWS96 combine matter
  bounded (MB) and ionization bounded (IB) components so that the MB
  component absorbs $\sim40$\% of the ionizing spectrum, corresponding
  to the exhaustion of He$^{+}$ ionizing photons and the peak in the
  effective heating rate in the cloud. The IB component is illuminated
  by the ionizing continuum that has been filtered by the MB
  component.  The MB component is responsible for most of the
  \ion{He}{2} emission, and the ionization bounded component emits low
  to intermediate excitation lines.  Previous multi-component models
  considered smooth distributions of cloud sizes (\cite{mor91} 1991)
  and combinations of thin MB clouds with IB clouds (\cite{vie92}
  1992).  In BWS96 IB and MB components are combined into a sequence
  parameterized by $A_{M/I}$, the solid angle ratio of MB clouds
  relative to IB clouds.
  
  In computing the photoionization models we have used the
  multi-purpose MAPPINGS II code. This is the same code as was
  used for the DS96 shock models, and is most recently described
  in \cite{sut93} (1993). The same solar abundance set as used by 
  DS96 was used for photoionization models. 
  This gives the advantage of removing abundance and atomic data
  differences and facilitates a direct comparison between shock and
  photoionization models.  We compute a set of $U$ sequences using a
  power-law ionizing source input spectrum ($F_\nu\propto\nu^\alpha$)
  with spectral indices $\alpha=-1,-1.4$. A high energy cutoff of 100
  Ryd (1.36 keV) was chosen so as to avoid over predicting the
  intensity of the soft X-rays with such flat spectral indices. The
  range in ionization parameter was $10^{-4} < U < 1$, and density
  $100< n <1000\ $cm$^{-3}$.  The models were ionization bounded
  with a matter filling factor of unity, and correspond to a range in
  cloud sizes 0.003$-$32 pc.

  The line ratios for the $A_{M/I}$ sequence are derived from the line
  intensities for MB and IB components given by BWS96. Table (2) in
  BWS96 gives the intensities for the MB and IB components for the
  single case of an $\alpha=-1.3, U_{MB}=0.04$, power-law continuum
  incident on the MB clouds, and the IB clouds illuminated by the
  continuum that is filtered by the MB clouds.  These parameters were
  chosen by BWS96 so as to give an optimum description of Seyfert and
  LINER spectra.  Line ratios for a given $A_{M/I}$ were calculated
  according to the prescription given by equation (2) in BWS96.  We
  consider the range $0.001< A_{M/I} < 100$ which encompasses the
  range used in the BWS96 ( $0.04< A_{M/I} < 16$). A larger value of
  $A_{M/I}$ gives a larger weight to the MB component, and therefore a
  higher excitation spectrum. Also, as discussed in BWS96, $A_{M/I}$
  cannot strictly be less than unity because the MB clouds are by
  definition intervening between the IB clouds and the ionizing
  source. An apparent $A_{M/I}<1$ may however be observed if some of
  the MB clouds are obscured along the line of sight.  Variations on
  this sequence are clearly possible, for instance by varying
  $U_{MB}$, $\alpha$, or the thickness of the MB component. However
  the $A_{M/I}$ sequence used in this paper is the only such sequence
  currently available in the literature and further investigation of
  the potential of this approach is beyond the scope of this paper.

\subsection{Shock models}

  Shocks have been incorporated into models of the NLR in various
  ways. In the simplest picture shocks serve to compress the
  gas which is then illuminated by a central ionizing source
  (\cite{ped85} 1985; \cite{tay92} 1992). This situation is then
  modelled with simple photoionization codes.  \cite{vie89} (1989)
  considered composite shock and photoionization models for clouds
  flowing either into, or away from, the nucleus. Models for which the
  shock itself generates enough ionizing radiation to form an
  equilibrium \ion{H}{2} region in the precursor gas, called
  photoionizing shocks\footnote{The name ``autoionizing shocks'' used
  by some authors is avoided because of the confusion with the atomic
  process of autoionization that has a separate meaning} have been
  presented by DS96 form the basis for the shock comparisons discussed
  in this paper.  DS95 present the line ratios predicted by these
  models, for a wide range of input shock parameters, as grids on
  optical line ratio diagnostic diagrams. They suggest that this
  mechanism may be solely responsible for the ionization of the NLR.

  The physics of photoionizing shocks is described in detail in
  DS96. Their simplified, 1-d steady-flow, radiative shock models show
  that a fast radiative shock generates a strong local radiation field
  of ionizing photons in the high temperature, radiative zone behind
  the shock front. The dominant emission processes here are electron
  thermal bremsstrahlung and line emission. This radiation field is
  isotropic, and affects the ionization and thermal balance of the gas
  both ahead of, and behind, the shock front.  Photons diffusing
  upstream may form an extensive precursor \ion{H}{2} region.  Photons
  diffusing downstream influence the ionization and temperature
  structure of the recombination zone of the shock. Transverse
  magnetic fields also play an important role by contributing to the
  pressure in the post-shock gas thereby limiting the compression
  factor of the shock.  The collisionally excited UV lines are
  predominantly produced in the hot cooling zone of the shock close to
  the recombination region ($10^5 >T_e ({\rm K})>10^4$).  The hotter
  post-shock region (T$_e\sim10^6$ K) is the source of the harder
  ionizing radiation generated in the shock.
  
  The shock model grids have been calculated in the low density limit
  \footnote{Densities $n_e<1000$\ cm$^{-3}$ in the photo-absorption zone,
   corresponding to $n_e\lesssim10$ for the preshock density. This
   ensures that lines used for diagnostic purposes are not seriously
   affected by collisional de-excitation.},
  and are parameterized by the shock velocity $V_{shock}$, and the
  magnetic parameter $B/\sqrt{n}$. The shock velocity controls the
  shape of the ionizing spectrum produced by the shock, and the
  magnetic parameter controls the effective ionization parameter in
  the optically emitting recombination zone of the shock. The output
  line spectra from these models are available as part of the AAS
  CD-ROM Series, Vol. 7 (\cite{lei96} 1996). DS96 give scaling
  relations for the total bolometric, and \Hb\ luminosities emitted by
  a shock, and show via optical line ratio diagrams that the output
  spectra from the shock models are similar to AGN spectra.

  In section 3 we plot the DS96 model grids for combinations of these
  UV and optical lines. Both shock+precursor, and shock-only
  grids are included. The range in shock velocity is $150 < V_{shock}
  < 500$ \kms\ for the shock only grids and $200 < V_{shock} < 500$
  \kms\ for the shock+precursor grids.  The range of magnetic
  parameter is $0 < B/\sqrt{n} < 4$\ $\mu$G\ cm$^{3/2}$.  Values of
  $B/\sqrt{n}\sim2-4$\ $\mu$G cm$^{3/2}$ correspond to equipartition
  of the thermal and magnetic pressures in the pre-shock plasma.  The
  intensity of \Heii\ is not given in the DS96 models, so we take the
  \Heii\ line flux to be 9 times the flux of \He4686.

\subsection{Limitations of the models}

  The most serious limitation of the shock models is the assumption of
  a steady one-dimensional flow. Secondary shocks which may occur in a
  proper three-dimensional treatment would affect the predictions of
  the lower ionization species most strongly.  Another assumption is
  that the radiative transfer is treated in the downstream-only
  approximation.  The transfer of the resonance lines, such as
  Ly$\alpha$ and \Civ, is treated according to a simple escape
  probability formulation for the local slab.

  Resonance line transfer is, of course, highly sensitive to geometrical
  factors.  In a planar shock the optical depth for resonance
  scattering will be highest in the compressed gas along the direction
  perpendicular to the shock front, so that the resonance line photons
  will be scattered preferentially into the upstream and downstream
  regions.  The resonance line emission from the shock is then
  expected to be dependent on viewing angle in the sense that a shock
  viewed side-on will have lower relative flux of resonance lines.

  Geometrical considerations are also important for photoionized gas.
  \cite{vil96} (1996) have considered the effects of resonance
  scattering and dust on the UV line spectrum of radio
  galaxies. Ly$\alpha$ is the most affected line, because it is
  subject to resonance scattering with very high optical depth both
  inside and outside of the ionized region, and is also absorbed by
  dust in regions of neutral material. We do not use Ly$\alpha$ in any
  of our diagnostics because these complications destroy any
  predictive capacity of this line. The effect on \Civ\ is not as
  severe because the regions for resonance scattering of this line are
  limited to zones where C$^{3+}$ exists, and in any case the line
  optical depths are far smaller than the case of Ly$\alpha$.
              
\section{DIAGNOSTIC DIAGRAMS}  

\subsection{Utility of UV diagnostics}

  As described in the introduction, it is difficult to discriminate
  between fast shocks and photoionization by a hard photon spectrum,
  with the standard optical diagnostic diagrams. In shocks faster than
  200 km\ s$^{-1}$, the photoionized precursor dominates the emission
  of many lines in the spectrum. Furthermore because the ionizing
  radiation responsible for the precursor has a substantial component
  of EUV photons it generates optical spectra which are not unlike
  those produced by a power-law with an upper cutoff in
  frequency. Lower velocity shocks, or shocks without precursors
  (which have been the main type of shock models reported in the
  literature) generate optical line ratios which are almost
  indistinguishable from power-law photoionization with a very low
  ionization parameter ($U<10^{-3}$).

  A number of authors have searched for specific line ratios which are
  able to separate shocks and photoionization. \cite{dia85} (1985) and
  \cite{kir89} (1989) discussed the use of the ratio \Sii\ to the near
  infrared lines \Siii\ to distinguish between shock and
  photoionization mechanisms in LINERs, while \cite{ost92} (1992)
  considered other near infrared lines. \cite{kir89} (1989) found that
  the shock models of \cite{bin84} (1984) were separated from the
  power-law photoionization models of \cite{sta84} (1984) on the
  \Sii/\Ha\ $vs.$ \Siii/\Ha\ diagram, and on other diagrams which
  employ the \Sii/\Siii\ ratio. \cite{kir89} (1989) found also that
  the observed line ratios of AGN fell in between the two sets of
  models, so a clear distinction of the excitation mechanisms was not
  possible in this case.  In that data set the LINERs showed the most
  scatter with respect to the models, while the Seyferts were more
  tightly grouped and fell closer to the photoionization models but
  still had too low \Siii\ compared to the models.

  DS96 present their shock models on a set of diagnostic diagrams,
  along with a database of Seyfert and LINER spectra. They found that
  Seyferts lie closer to predictions of shocks with precursors, while
  LINERs have line ratios predicted by shocks without precursors
  although there is significant scatter of the observations around the
  model grids. The ambiguity of determining the excitation mechanism
  from these lines arises because photoionization models have a
  considerable amount of overlap with both the shock and
  shock+precursor grids. DS96 emphasize the use of ultraviolet
  emission lines as better discriminants between shocks and
  photoionization because the UV resonance lines (eg. \Civ) which are
  predominantly produced in the cooling zone of the shock are
  predicted to be much stronger than in photoionization models that
  give the same optical spectrum.

  Since the predictions of particular UV line ratios differ
  significantly between shock and photoionization models, it is
  possible to construct UV, and UV-optical line ratio diagrams in
  which the model grids are separated. In order to make these diagrams
  a practical tool for discriminating between the models we adopt
  criteria similar to \cite{vei87} (1987). We select lines which are
  relatively bright in active galaxies, and where possible use ratios of 
  lines that are relatively close in
  wavelength to minimise the sensitivity to reddening corrections.

  The line ratios we find most useful for comparing observations with
  the models are the UV line ratios \Civ/\Heii, \Civ/\Ciii,
  \Cii/\Ciii, the near UV lines \Nev/\Neiii. The ratios involving
  FUV lines \Niii/\N991 and \Ciii/\C977, while not close in
  wavelength are included as a temperature sensitive diagnostic.  In
  constructing line ratio diagrams we find that it is useful to
  combine these ratios with the optical ratio \Oiii/\Hb, in order to
  better separate the various model grids.  The diagnostic diagrams
  are presented in Figures 2 and 3 and we discuss each diagram
  individually in section 3.3. We also compare the new diagrams with
  one of the standard optical line ratio diagrams \Oiii/\Hb\ $vs.$
  \Nii/\Ha.

  For high redshift objects the UV lines are shifted into the optical
  band and the rest wavelength optical lines are less accessible,
  being in the near infrared. For these objects it is more useful to
  have diagnostics that rely only on UV lines. \cite{vil97} (1997;
  hereafter VTC97) have recently used line ratio diagrams involving
  \Civ/\Heii, \Ciii/\Heii\ and \Civ/\Ciii\ to compare a database
  of 21 high-$z$ radio galaxy spectra ($z>1.7$), with various models
  including the shock models from DS96. On these diagrams the line
  ratios for the precursor gas have been plotted separately, rather
  than having the shock and precursor spectra summed to form a
  shock+precursor grid. Since the precursor gas is not resolved from
  the shocked gas it is inappropriate to compare such data with the
  precursor-only line ratios as the emission from the shock will also
  contribute to the observed spectrum. In section 3.5 we repeat the
  diagrams used by VTC97, with the shock-only, and shock+precursor
  grids appropriately combined.

\subsection{Comparison with the observations}

  It is not the purpose of this paper to present a large data base of
  observations for comparison with the models. Instead, for most of
  our diagrams which use UV and optical lines, we confine the
  observational data to a small sample of objects for most of which we
  have obtained long wavelength baseline spectra with the $HST$ FOS
  and ground based telescopes.  We also include combined IUE, and HUT
  data on NGC 1068 collected from the literature. Since the
  diagnostics presented below rely on ratios of both optical and UV
  lines, we emphasize the need for wide wavelength coverage spectra to
  obtain as many of these diagnostic lines as possible.

  For high-$z$ objects the observational material has only UV coverage
  so we are obliged to use diagrams involving UV lines alone. However,
  the average high-$z$ spectrum given by \cite{mcc93} (1993) has a
  wider spectral coverage, and can therefore be included in most
  diagrams.

  Spectra of the NLRs in the Seyfert 2 galaxies NGC~5643 and NGC~5728
  have been obtained with the FOS using the 0\farcs86 square
  aperture. Both NGC~5643 and NGC~5728 have striking ionization cones
  seen in HST narrow band images (\cite{wil93} 1993; \cite{sim97}
  1997) and show evidence for hidden nuclei. The aperture for the
  observations was located over the bright apex of the cone in each
  case. The high resolution FOS spectra covered the wavelength range
  $1150-2500$ \AA, and observations with the FOS prism covering
  $1150-5400$ \AA\ allowed us to tie the flux scale to ground based
  optical observations.  Ground based spectra were taken with the ANU
  2.3 m telescope at Siding Spring with wavelength coverage
  $3000-10000$ \AA. The details of the FOS spectra and the procedure
  used for joining the FOS and ground based spectra will be presented
  in a future paper.
  
  The FOS spectrum of the central disk in M87 was taken with the
  0\farcs86 circular aperture positioned 0\farcs6 from the nucleus,
  avoiding contamination of the emission line spectrum by continuum
  from the nucleus. The wavelength coverage is $1150-7000$ \AA\ and
  the spectrum reveals LINER-like line ratios. A full line list is
  given in \cite{dop97} (1997) along with a detailed study of the
  central disk.

  The line ratios for NGC~5643, NGC~5728 have been corrected for
  reddening assuming an intrinsic \Ha/\Hb\ ratio of 3.1, and using the
  reddening function of \cite{car89} (1989). The corresponding visual
  extinctions for our NGC~5643 and NGC~5728 spectra are A$_V$ = 1.08,
  and A$_V$ = 1.26 respectively. The reddening for the M87 spectrum is
  taken to be A$_V$=0.124 (\cite{dop97} 1997).

  For NGC~1068, the very well studied Seyfert 2 galaxy, we have used
  the collated line flux measurements drawn from the literature by
  \cite{dop97a} (1997). This is the only object for which we have
  measurements of the FUV lines \N991\ and \C977, these come from the
  Hopkins Ultraviolet Telescope (HUT; \cite{kri92} 1992).

  The database of high-$z$ radio galaxies is taken directly from Table
  1 in VTC97. This consists of \Civ/\Ciii, \Civ/\Heii\ and
  \Ciii/\Heii\ measurements for 21 radio galaxies with $z>1.7$,
  mostly from \cite{oji95} (1995), plus single objects from
  \cite{els94} (1994), \cite{mcc90} (1990), and \cite{spi85} (1985).

\subsection{UV diagnostic diagrams}

  In all the diagrams presented below the shock-only, shock+precursor,
  and photoionization model grids are plotted with consistent line
  type and grid spacing. The shock-only grids are plotted in grey, and
  are labelled with magnetic parameter $B/\sqrt{n}$= 0, 1, 2, 4 $\mu$G
  cm$^{3/2}$, and shock velocity $V_{shock} = 150$, 200, 300, 500 km\
  s$^{-1}$. The shock+precursor grids are labelled with $V_{shock} = 
  200$, 300, 500 km\ s$^{-1}$, and magnetic parameter where possible.  
  The $\alpha=-1$ and the $\alpha=-1.4$ power-law photoionization $U$ 
  sequences are plotted for $n$ = 100 cm$^{-3}$ and $n$ = 1000 cm$^{-3}$ 
  with grid lines of constant $U$ every 0.25 dex. The $A_{M/I}$ 
  sequence is labelled with $A_{M/I}$ over the range $0.001 
  \leq A_{M/I} \leq 100$, and tick-marks at intervals of 0.2 dex.

  The reddening corrected data points for M87, NGC~1068, NGC~5643 and 
  NGC~5728 are plotted
  as solid circles with their error bars, and the object names are
  indicated next to the points. Where possible we have included a data
  point (star) for the average high redshift radio galaxy of
  \cite{mcc93} (1993).

  All diagrams include a reddening vector which indicates the effect
  of an extinction of 10 $A_V$. Note that due to the 2175 \AA\ bump in
  the extinction law, the line ratio \Civ/\Ciii\ decreases rather
  than increases with extinction.

  Figure 1, \Oiii/\Hb\ $vs$. \Nii/\Ha, is one of the standard optical
  diagnostic diagrams. It is included here for comparison with the UV
  and UV-optical diagrams. The shape of the shock and shock+precursor
  grids on this diagram have been discussed by DS95 where the models
  are compared with a database of Seyfert and LINER spectra. DS95
  finds that LINER spectra are closer to the shock-only grid while the
  Seyferts have spectra closer to the shock+precursor
  predictions. Power law photoionization models can also generate
  spectra in the same region of the diagram as the shock-only and
  shock+precursor model grids.  In Figure 1 the high velocity part of
  the shock-only grid coincides with the power-law photoionization
  sequences for $U \sim10^{-3.5}$, and the high velocity end of the
  shock+precursor grid overlaps the higher $U \sim10^{-2.5}$
  photoionization models. In the following optical-UV and UV
  diagnostic diagrams we demonstrate that the pure shocks can be
  cleanly separated from the photoionization, and that $V_{shock}<400$
  km\ s$^{-1}$ shocks with precursors can be discriminated from
  photoionization.

\subsubsection{ (\Oiii/\Hb) vs. (\Civ/\Heii) and (\Civ/\Ciii) } 
  
  In Figures 2a and 2b we combine the commonly used \Oiii/\Hb\ ratio
  with the UV ratios \Civ/\Heii, and \Civ/\Ciii. The UV ratios have
  been chosen with the criteria discussed above, and specifically to
  include \Civ\ because of the large differences in the shock and
  photoionization model predictions for the strength of this line.  We
  note that \Civ/\Heii\ utilizes \Heii\ as a recombination line
  reference in the UV, much as \Hb\ is used in the optical. The
  \Civ/\Ciii\ ratio has the additional advantage of being independent
  of abundance.

  In shocks, the \Heii\ line flux relative to the Balmer lines
  increases monotonically with shock velocity (DS95) and \Civ\ is
  always strong as a result of collisional excitation. In contrast,
  photoionization models generate similar \Heii\ flux for a given
  spectral index since this line arises from recombination, and the
  \Civ\ line intensity is determined primarily by the temperature
  structure of the \ion{H}{2} region, being generally stronger as the
  ionization parameter increases.

  It is useful to combine the UV line ratios \Civ/\Heii\ and
  \Civ/\Ciii\ with the optical ratio \Oiii/\Hb\ since this ratio
  separates the two types of shock models. Simple shock models without
  photoionized precursors have lower \Oiii/\Hb\ than shock+precursor
  models because the \Oiii\ emission is dominated by the precursor
  zone.  The ionization state of the precursor gas is a strong
  function of shock velocity, so in the shock+precursor case,
  \Oiii/\Hb\ increases monotonically with shock velocity, from 1.5 at
  $V_{shock}$ = 200 km\ s$^{-1}$ to 12.6 at $V_{shock}$ = 500 km\
  s$^{-1}$. The dependence on shock velocity arises because higher
  velocity shocks produce a harder spectrum of ionizing photons, which
  increases the temperature and effective ionization parameter in the
  precursor zone.  Under these conditions \Oiii\ becomes relatively
  more efficient as a coolant, with the effect of increasing
  \Oiii/\Hb. Figures 2a and 2b show the separation of shock+precursor,
  and shock-only grids along the \Oiii/\Hb\ axis. The two grids begin
  to merge for low velocity shocks where the precursor emission is
  negligible, and the line ratios are dominated by the shock rather
  than by the precursor.
  
  Figures 2a and 2b confirm that \Civ/\Heii\ and \Civ/\Ciii\ are
  strong functions of $U$ in the power-law photoionization models.
  For example \Civ/\Heii\ increases from 10$^{-4.5}$ at $U=10^{-4}$,
  to approximately 1.6 at $U=0.1$. The photoionization model grids
  show that in order to produce the \Civ/\Heii\ ratio as strong as
  predicted by the shock models (0.4$-$16), a relatively high
  ionization parameter $U\gtrsim0.01$ is required.  The ionization
  parameter required to reproduce the UV line strengths observed in
  AGN is generally higher than that indicated by the observed optical
  lines (cf. Figure~1). Power-law photoionization models for Seyferts
  typically require ionization parameters $U=10^{-3}$ to 10$^{-2}$
  ($\alpha\sim-1.5$) to reproduce the optical line ratios
  (\cite{ho93b} 1993).  LINERs, however, have low excitation optical
  spectra, with \Oiii/\Hb$\lesssim3$. Photoionization models for
  LINERs typically adopt a similar ionizing continuum to that used for
  Seyferts, $\alpha\sim-1.5$, but require a much lower ionization
  parameter ($U\sim10^{-3.5}$) to reproduce the optical line ratios
  (\cite{ho93} 1993). This can be seen in Figure 1, where the Seyferts
  NGC~1068, NGC~5643 and NGC~5728 indicate $U\sim10^{-2.5}-10^{-2.8}$
  with respect to the photoionization models.  The single zone, low
  ionization parameter photoionization models are however, unable to
  account for the strong UV lines and the \Civ/\Heii\ ratio observed
  in some LINERs, such as M87. This is because, when the ionization
  parameter is low enough to produce a LINER-like spectrum in the
  optical, the electron temperature is too low to give appreciable
  excitation of the \Civ\ line, or any other line in the far-UV.

  Multiple-component photoionization models go some way to solve the
  problems of single zone models. The two component models of BWS96
  which combine a MB component and an IB component, are able to
  produce relatively strong UV lines in combination with a low
  excitation optical spectrum. The $A_{M/I}$ sequence shown in
  Figure~2a has values of \Civ/\Heii\ in the same range as the shock
  models. For values of $0.001<A_{M/I}<0.05$ the sequence is dominated
  by the IB component, with only a small fraction of MB clouds.  Over
  this range of $A_{M/I}$ the \Oiii/\Hb\ ratio remains relatively low
  at approximately 3.5. This value is comparable to the value of
  \Oiii/\Hb\ predicted by power-law models with $U<0.001$, although
  the input ionizing spectrum incident on the MB clouds has $U_{\rm
  MB}$=0.04.
 
  It is important to note that, while low values of $A_{M/I}$
  correspond to a lower proportion of MB clouds, and a higher
  proportion of IB clouds, the line ratios calculated for the low
  $A_{M/I}$ end of the sequence do not generally converge to the
  values produced by the ionization bounded $U$ sequence of the same
  ionization parameter. This is because the ionizing continuum which
  illuminates the IB component of the $A_{M/I}$ sequence is still
  filtered by the MB component even for low values of $A_{M/I}$, and
  is not therefore the same as the power law continuum used in the $U$
  sequences. As was mentioned in section 2.1, values of $A_{M/I}$ less
  than unity must correspond to an apparent $A_{M/I}$, in which the MB
  clouds are obscured from direct viewing along the line of sight.
    
  When the ionization parameter for the photoionization models exceeds
  $\sim0.1$ the ionization structure of the cloud changes so that
  ionization balance for Carbon shifts to higher ionization stages,
  decreasing the emission of \Civ\ relative to \Heii.  This effect
  causes the photoionization $U$ sequence curves to reach a maximum in
  \Civ/\Heii\ at 3.2 for the $\alpha=-1$ models and 1.6 for the
  $\alpha=-1.4$ models. The $U$ sequences then fold over themselves at
  the high $U$ end of the sequence, so that for some regions in Figure
  1 there is a double valued photoionization solution. The diagnostic
  capability of this region of the diagram is limited by this effect
  which prevents the photoionization parameters from being uniquely
  determined, but the diagnostic capability is more seriously
  undermined by the fact that the photoionization models overlap with
  shock+precursor models with high magnetic parameter, and shock
  velocities in excess of 400 km\ s$^{-1}$. The reason for the overlap
  is that the ionizing radiation produced by a fast shock is not
  unlike a power-law distribution, so produces similar spectra to the
  high $U$ photoionization models.
 
  In the photoionization models the ratio \Civ/\Ciii\ behaves in a
  similar way to the \Civ/\Heii\ ratio for values of the ionization
  parameter $U\lesssim0.1$. For higher values of $U$, the \Civ/\Ciii\
  ratio continues to increase whereas the \Civ/\Heii\ ratio reaches a
  maximum. The photoionization grids in Figure~2 overlap with the high
  shock velocity ($V_{shock} > 400$ km\ s$^{-1}$), shock+precursor
  grids, showing that photoionization models can produce values of the
  \Civ/\Ciii\ ratio which lie in the same range as shocks, but in
  order to do so, a high ionization parameter is required. The
  $A_{M/I}$ sequence in Figure~2 has values of \Civ/\Ciii\ in the
  range predicted by shocks for values of $A_{M/I}>0.1$. Below this
  value, the \Civ/\Ciii\ ratio decreases rapidly with $A_{M/I}$.

  Comparing our set of observations to the model grids in Figure~2a,
  we find that the NLRs of the Seyfert 2 galaxies NGC~1068, NGC~5643,
  and NGC~5728 have line ratios which fall in a region where the
  shock+precursor grid, the high $U$ end of the power-law
  photoionization $U$ sequences, and the A$_{M/I}$ sequence all
  overlap. This obviously prevents us from unambiguously separating
  the emission mechanisms with these particular line ratios.
  
  M87 has a LINER-like optical spectrum, and its low \Oiii/\Hb\ ratio
  clearly separates it from the Seyferts in our diagrams.  In both
  Figures 2a and 2b, M87 falls closest to the shock-only grid, with
  shock velocities between 200 and 300 km\ s$^{-1}$, and high magnetic
  parameter. This is a good example of the power of the UV diagnostics
  over the optical diagrams. For example, in Figure 1 objects with
  LINER like spectra such as M87 lie close to both shock-only and $U
  \sim10^{-3.5}$ photoionization models, whereas Figures 2a and 2b
  which include UV lines, cleanly separate photoionization and
  shock-only grids which allows us to identify the M87 spectrum as
  shock excited.  Single zone photoionization models for M87 have been
  explored by \cite{dop97} (1997), where power-law, blackbody, and
  Bremsstrahlung continua have all been tuned to reproduce the
  observed \Oiii/\Hb\ ratio. None of these cases can reproduce the UV
  lines which were predicted to be much weaker than
  observed. Furthermore, by using the shock grid as a starting point,
  \cite{dop97} (1997) have calculated an optimized shock model for M87
  giving a self-consistent shock velocity of 265 km\ s$^{-1}$, and
  magnetic parameter B/$\sqrt{n}=5.4\ \mu$G\ cm$^{3/2}$.
  
\subsubsection{ (\Oiii/\Hb) vs.\ (\Nev/\Neiii)}
 
  \Nev\ is another high ionization line which, like \Civ, is always
  strong in shocks whilst being strongly dependent on ionization
  parameter in power-law photoionization models. We utilize the ratio
  of \Nev\ to \Neiii, these lines are close in wavelength, and the
  ratio is abundance independent.  The shapes of the photoionization and
  shock model grids in Figure~2c are quite similar to Figures 2a and
  2b. Like \Civ/\Heii\ in Figure 2a, \Nev/\Neiii\ reaches a maximum in
  the photoionization models near $U=0.1$, but in contrast to Figures
  2a and 2b, the $A_{M/I}$ sequence produces the highest excitation
  values of \Nev/\Neiii.
  
  On this diagram, Figure~2c, the Seyferts NGC~5728 and NGC~5643 fall
  within the $\alpha=-1.4$ power-law models, in the region where the
  grids fold back on themselves. Although this overlap prevents an
  accurate determination of the ionization parameter with these
  ratios, the points indicate ionization parameters in the range
  0.03$-$0.3.  NGC~1068 falls just outside the photoionization grids
  closer to the $A_{M/I}$ sequence.  Compared to the shock+precursor
  models the Seyfert observations indicate an \Nev/\Neiii\ ratio
  approximately 0.5 dex higher than the shock+precursor grid.  The M87
  point is again consistent with the shock-only models, falling in the
  same location with respect to the grid as in Figure~2b.

  A practical advantage of using \Nev\ and \Neiii\ with \Oiii\ and
  \Hb\ is that these lines are accessible with ground based
  instruments for low redshift objects.

\subsubsection{ (\Cii/\Ciii) vs. (\Civ/\Ciii)}
 
  We find that the combination of the line ratio pairs \Cii/\Ciii\
  and \Civ/\Ciii, Figure~2d, are able to separate the
  photoionization model grids from the shock-only grid, and are also
  able to separate the shock+precursor grid from the photoionization
  grids to a greater extent than the previous diagrams.  In this
  respect we regard this to be an important UV diagnostic for
  discriminating between the two competing mechanisms.  Also, since
  the ratios involve only lines of different ionization stages of
  Carbon it is also insensitive to abundance effects.

  In order to provide a unique discriminant between models on line
  ratio diagrams, model grids must not overlap.  While there is
  formally no overlap between the shock+precursor grid and the
  photoionization $U$ sequences, the $B/\sqrt{n}=0\ \mu$G\ cm$^{3/2}$,
  $V_{shock}=500$ km\ s$^{-1}$ corner of the shock+precursor grid does
  approach very close to the $\alpha=-1$ $U$ sequences. Note however
  that the $B/\sqrt{n}=4\ \mu$G\ cm$^{3/2}$, $V_{shock}=500$ km\
  s$^{-1}$ region of the shock+precursor grid, which overlaps the
  $\alpha=-1.4$ photoionization models in Figures 2a, 2b and 2c is
  separated from the photoionization models in Figure~2d. The
  $A_{M/I}$ sequence crosses the $U$ sequences at both the low and
  high end of the $A_{M/I}$ parameter values, but otherwise has a
  distinct locus.  Also, part of the $A_{M/I}$ sequence overlaps with
  the high velocity end of the shock+precursor grid, as it does in
  Figures 2a and 2b.

  The NGC~1068 data point falls at the same position with respect to
  the shock+precursor grid in Figure~2d, as it did in Figures~2a and
  2b, indicating that the $B/\sqrt{n}=4\ \mu$G\ cm$^{3/2}$,
  $V_{shock}=480$ km\ s$^{-1}$ shock+precursor model correctly
  reproduces the observed \Cii, \Ciii, \Civ, \Heii, \Oiii\ and \Hb\
  fluxes for NGC~1068. The NGC~5643 and NGC~5728 points in Figure~2d
  fall inbetween the shock+precursor grid and the $\alpha=-1$
  photoionization grids, making it difficult to assign either
  mechanism as dominant. NGC~1068 falls close to the $A_{M/I}$
  sequence for $A_{M/I}\sim0.5$, which is lower than indicated in
  Figures 2a, 2b and 2c, where the sequence indicates $A_{M/I}>1$.

  Because the shock-only model grids are well separated from the
  photoionization grids with the line ratios used in Figure 2 the
  \Cii/\Ciii\ $vs.$ \Civ/\Ciii\ diagram constitutes a powerful
  diagnostic for discriminating between shock and photoionization
  mechanisms for LINER-like spectra. We find that M87 is a good
  example of this, whereby the optical line ratios can be equally well
  produced by photoionization or shocks, but that the UV ratios enable
  us to state unequivocally that this object is shock-excited.
  Consistent with all the previous diagrams, the M87 point falls on
  the shock-only grid with a shock velocity of $\sim$300 km s$^{-1}$
  and high magnetic parameter.

  As described above, the apertures used in the UV observations of the
  Seyferts encompassed the bright apex of the ionization cones near
  the inferred position of the hidden nucleus. It is not yet known how
  the UV line ratios vary spatially over the extent of the NLRs in
  these objects.  The optical line ratios of these objects are known
  to vary throughout the NLR, and in some cases are suggestive of
  shock excitation.  Spatially resolved UV spectroscopy of the NLRs of
  AGN will shed much more light on the relative importance of
  photoionization and shocks.

\subsection{Temperature sensitive diagnostics}

 The temperature sensitive ratios such as \Niii/\N991, and \Ciii/
 \C977\ are potentially very useful for discriminating between shocks
 and photoionization.  In shock models the FUV lines \Niii, and \C977\
 are generated almost entirely within the shock structure, where the
 temperatures are much higher than possible in any photoionization
 model. In Figures 3a and 3b these ratios are combined with the
 optical ratio \Oiii/\Hb. As in Figures 2a, 2b and 2c the \Oiii/\Hb\
 ratio serves to separate the shock-only and shock+precursor grids.

 In Figures 3a and 3b both types of shock model are completely
 separated from the photoionization models making this a very good
 diagnostic for discriminating between the models.  Temperatures
 indicated by the \Niii/\N991, and \Ciii/\C977\ ratios are plotted
 along the top axis of the diagrams in intervals of 5000K up to
 100000K. These temperature axes were calculated assuming purely
 collisional excitation in the low density limit ($n_e<10^5$
 cm$^{-3}$) and using the collision strengths of \cite{duf78} (1978)
 for the \ion{C}{3}\ ratio and \cite{blu92} (1992) for the \ion{N}{3}\
 ratio.  The observations of NGC~1068 derived from HUT and IUE spectra
 indicate high temperatures; 47900 K from the \Niii/\N991\ ratio and
 26700 K from the \Ciii/\C977\ ratio. The line ratios predicted by
 the shock models are indicative of temperatures from 25000 K to
 $>$100000 K.

 The A$_{M/I}$ sequences shown in Figures 3a and 3b were calculated in
 the same manner as all the previous diagrams, with MB and IB
 components for the \C977\ and \N991\ lines provided by Binette
 (private communication). \N991\ and \C977\ may be underestimated in
 the A$_{M/I}$ sequences as those calculations (done with MAPPINGS Ic)
 did not consider the effects discussed by \cite{fer95} (1995).

 The NGC~1068 observation indicates a temperature higher than possible
 with either the power-law photoionization or the A$_{M/I}$
 sequence. It is more consistent with shocks in Figure 3a but falls
 inbetween the shock and photoionization model grids in Figure 3b.

\subsection{High-$z$ diagnostics}  

 Here we consider the UV diagnostics which were used by VTC97
 to investigate the balance between shocks and photoionization
 in high-$z$ radio galaxies. They compare spectra of 21 high-$z$ radio 
 galaxies ($z>1.7$) with photoionization models, and the shock
 models from DS96 using similar UV diagnostic diagrams to those
 presented in this paper. In their models they explore 
 the effect of viewing angle on the spectrum of photoionized
 clouds. They consider clouds where the illuminated face is
 viewed directly, and the case where clouds are viewed from the rear.
 They find that $\alpha=-1.5$ power-law photoionization models, which 
 reasonably reproduce the optical line ratios of low redshift radio
 galaxies, cannot explain the observed UV line ratios of high-$z$ 
 radio galaxies. Instead a harder ionizing continuum of $\alpha=-1.0$
 gives a good fit to their data, with the points falling in the region
 between predictions for front and rear viewed clouds.

 VTC97 also compared their data to the same DS96 shock models as
 described in this paper. In their diagrams (Figure 3 in VTC97) the
 emission line ratios for the precursor gas of the shock models were
 plotted separately, rather than summing the shock and precursor
 spectra to form a shock+precursor grid. As described in section 3.1,
 plotting the precursor line ratios separately is inappropriate
 because the precursor gas is not spatially resolved from the shock
 gas in these distant galaxies. We have therefore recomputed these
 same diagrams in Figure 4, with the properly combined shock+precursor
 grid, and the shock-only grid. We plot the data points from
 VTC97. Note that this data set does not include include the rest
 wavelength optical lines, or the \Cii\ line because these lines are
 redshifted outside the spectral coverage. As such we cannot use the
 more sensitive diagnostics presented in Figure 2.  For completeness
 we compare these points to our photoionization models, the A$_{M/I}$
 sequences (for a wider range of A$_{M/I}$ than used by VTC97), and
 our observations of nearby AGN. We find that our simple
 photoionization models behave in a similar way on these diagrams to
 the photoionization models presented in VTC97.
 
 In Figure 4 we find that the shock-only models cannot account for the
 observed high-$z$ galaxy line ratios. The shock+precursor models
 overlap approximately one quarter of the observed points however the
 grids exhibit somewhat complicated behaviour on these diagrams,
 having a number of regions with multiple solutions for the line
 ratios. The photoionization models appear to explain the trends in
 the data set by varying the ionization parameter. For example in
 Figures 4e and 4f the points fall along the photoionization models
 with ionization parameter in the range $10^{-2} < U < 10^{-1.5}$,
 whereas the shock+precursor model grid overlaps only a subset of the
 data points. As such, the shock+precursor models are not able to
 explain the trends in the data, in terms of shock parameters, and we
 confirm the conclusion of VTC97 that an $\alpha=-1$ power-law
 photoionization model provides a good fit to the line ratios.
 
 As high redshift galaxies may have abundances that are less than
 solar, we include an arrow indicating the shift of the model grids
 due to a depletion of 0.5 dex in the Carbon abundance. The effect of
 resonance scattering of \Civ\ is considered in detail by \cite{vil96}
 (1996) and a decreased abundance of Carbon ($\sim$0.2 dex), could
 explain the offset of the high-$z$ AGN relative to the nearby AGN.

 Comparing our data set of nearby AGN to the models on these diagrams,
 we find that the NGC~1068 point falls on the shock+precursor grids
 with shock parameters consistent with those inferred by Figures 2a,
 2b and 2d. The other Seyferts, NGC~5643 and NGC~5728 also fall close
 to the shock+precursor models in the same sense as the previous
 diagrams, but the models are not as well separated on these diagrams
 and the Seyferts also fall on the the $\alpha=-1$ photoionization
 grid with ionization parameters in the range $10^{-2.2}< U
 <10^{-1.9}$. Similarly the shock-only and photoionization model grids
 are degenerate in Figure 4, making it difficult to determine the
 excitation mechanism for LINER-like spectra using these ratios alone,
 and we emphasize again that the more sensitive diagnostics in Figure
 3 allow us to discriminate between pure shocks and photoionization.

\section{CONCLUSIONS}

  In this paper we have presented UV and UV-optical diagnostic
  diagrams that are able to discriminate between shock and
  photoionization predictions for the line ratios in the narrow
  emission line regions in AGN. These diagnostics rely on the
  differences between the model predictions of the high ionization
  lines for separating shock line ratios from photoionization. In
  general, the diagrams discussed here show that we can, to a large
  extent, separate photoionization from shocks with precursors if the
  shock velocity is less than 400 km s$^{-1}$. Above 400 km s$^{-1}$
  the shock+precursor grids overlap with the photoionization models.
  The most useful diagnostics utilize the \Cii/\Ciii, \Civ/\Ciii\
  ratio pair, and also the near UV \Nev/\Neiii\ ratio provides good
  diagnostic that can be easily measured with ground based
  instruments.  These UV diagnostics are expected to form a useful
  observational tool for comparison with UV spectra obtained with the
  FOS and particularly for spatially mapped UV spectra that will be
  obtained with STIS.  We also have demonstrated that the FUV lines
  \C977\ and \N991\ are sensitive temperature diagnostics which can
  discriminate between the models.

\acknowledgments

  The results in this paper are based on observations made with the
  NASA/ESA Hubble Space Telescope, obtained at the Space Telescope
  Science Institute, which is operated by the Association of
  Universities for Research in Astronomy, Inc., under NASA contract
  NAS5-26555. MD thanks the Australian Dept. of Industry, Science and
  Tourism (DIST) for support under a International Science and
  Technology (IS\&T) major grant, which facilitated publication costs
  and travel. MA acknowledges the support of a Studentship at the Johns
  Hopkins University (1995-6) through grants GO-4371.01, GO-5426.01 and
  GO-5926.02 and travel support from the DIST IS\&T major grant. ZT was
  supported in part through NASA grants NAG5-1630 and NAGW-4443 to the 
  Johns Hopkins University.

\clearpage

\begin{deluxetable}{llll}
\footnotesize \tablecaption{Observations \label{tbl-1}}
\tablewidth{0pt} 
\tablehead{\colhead{Object} & \colhead{Instrument} & \colhead{$\lambda$}} 
\startdata 
NGC~5643 & $HST$ FOS                       & \phm{$\sim$}1150 $-$ 2500    \nl 
         & MSSSO 2.3m telescope            & \phm{$\sim$}3000 $-$10000    \nl 
NGC~5728 & $HST$ FOS                       & \phm{$\sim$}1150 $-$ 2500    \nl 
         & MSSSO 2.3m telescope            & \phm{$\sim$}3000 $-$10000    \nl 
M87      & $HST$ FOS$^1$                   & \phm{$\sim$}1150 $-$ 6820    \nl 
NGC~1068 & various$^{2}$                   & \phm{$\sim$}\phn977 $-$ 9600 \nl
         & HUT$^3$                         & \phm{$\sim$}\phn830 $-$ 1860 \nl 
         & IUE$^4$                         & $\sim$1100 $-$ 3300 \nl 
\tablerefs{ (1) \cite{dop97}; (2) compiled in \cite{dop97a}; (3) \cite{kri92} (1992); (4) \cite{sni86} (1986) }
\enddata
\end{deluxetable} 

\clearpage

\clearpage

\begin{figure}[tbp]
\plotone{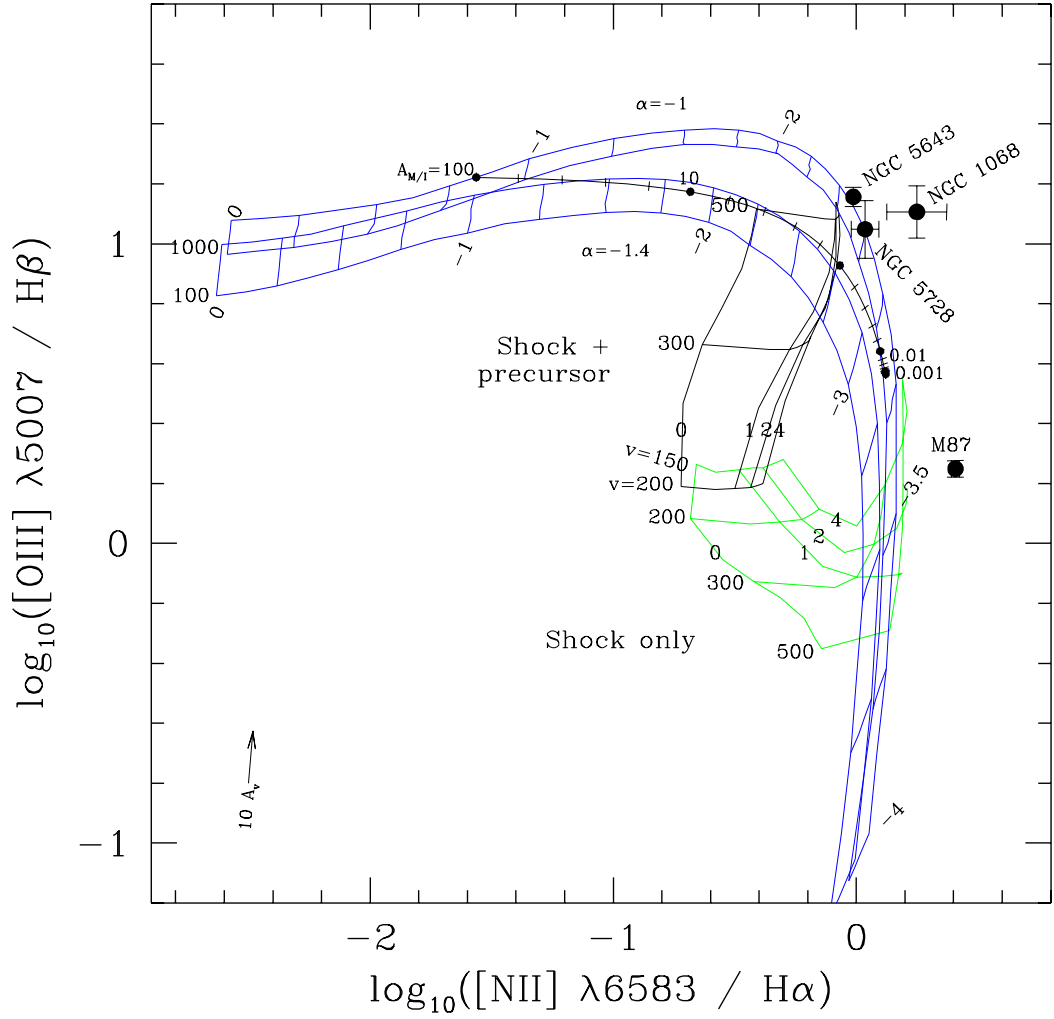}
\caption{Optical diagnostic diagram.}
\end{figure}
 
\clearpage
 
\begin{figure}[tbp]
\plottwo{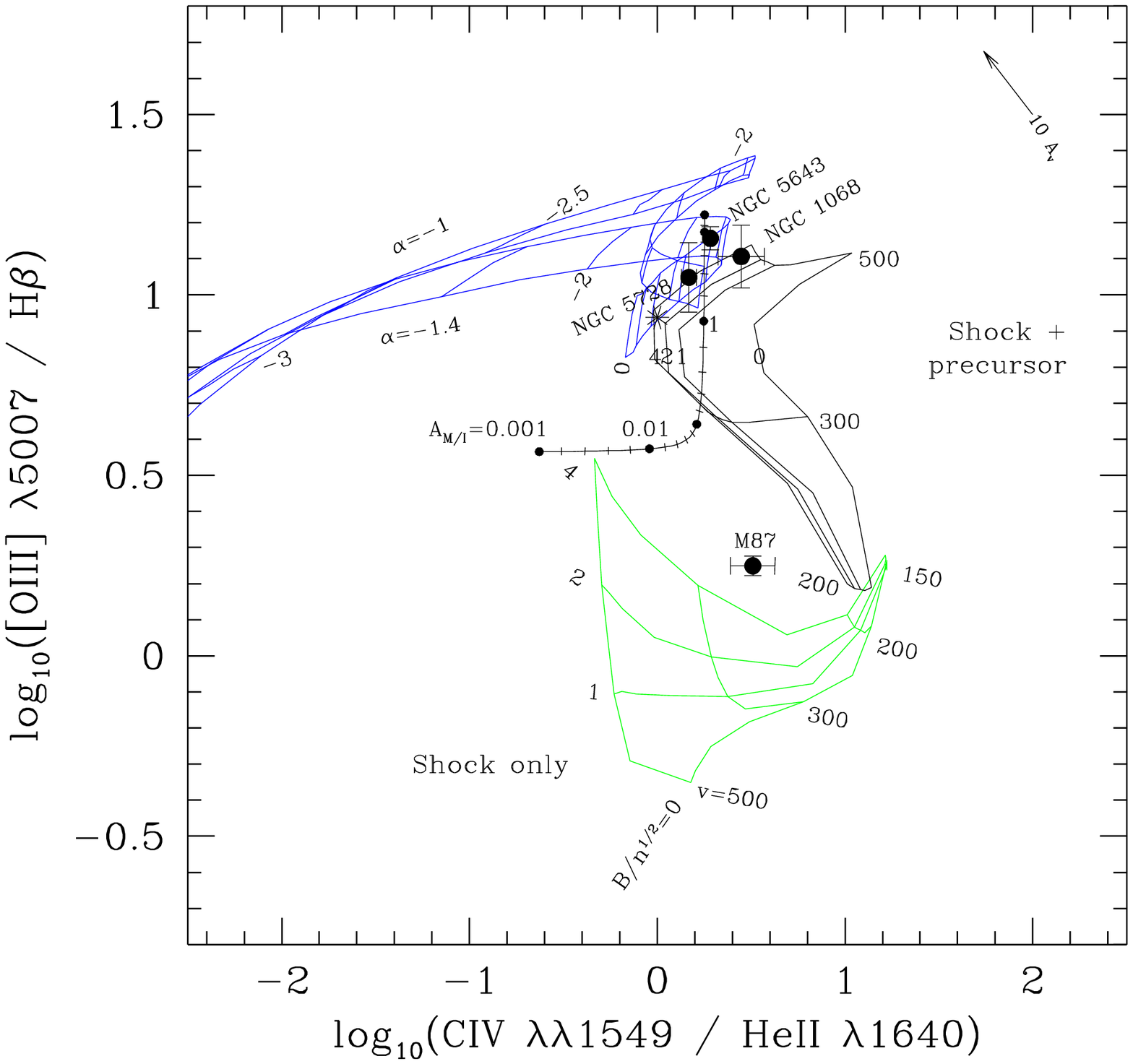}{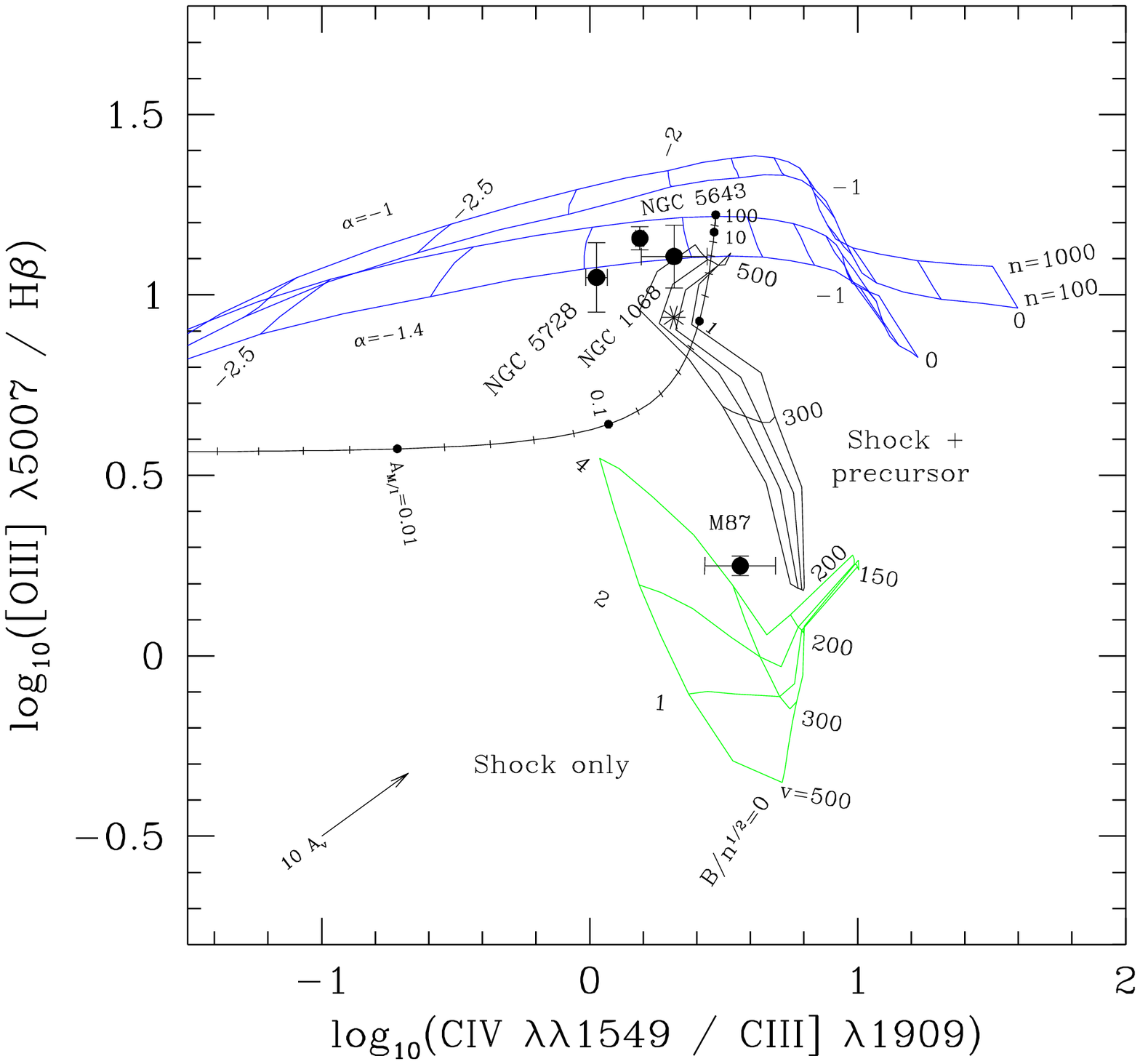} \newline
\plottwo{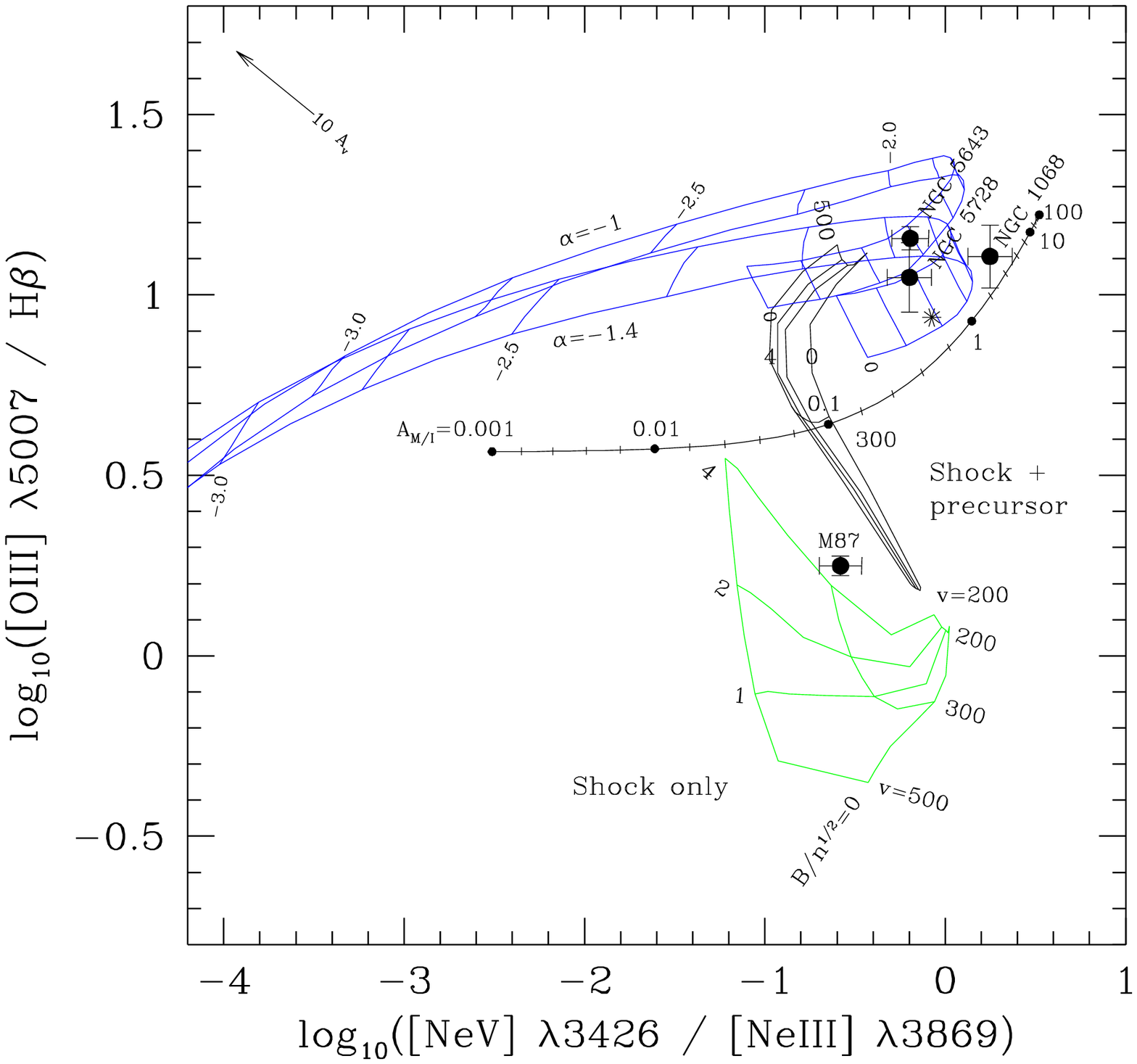}{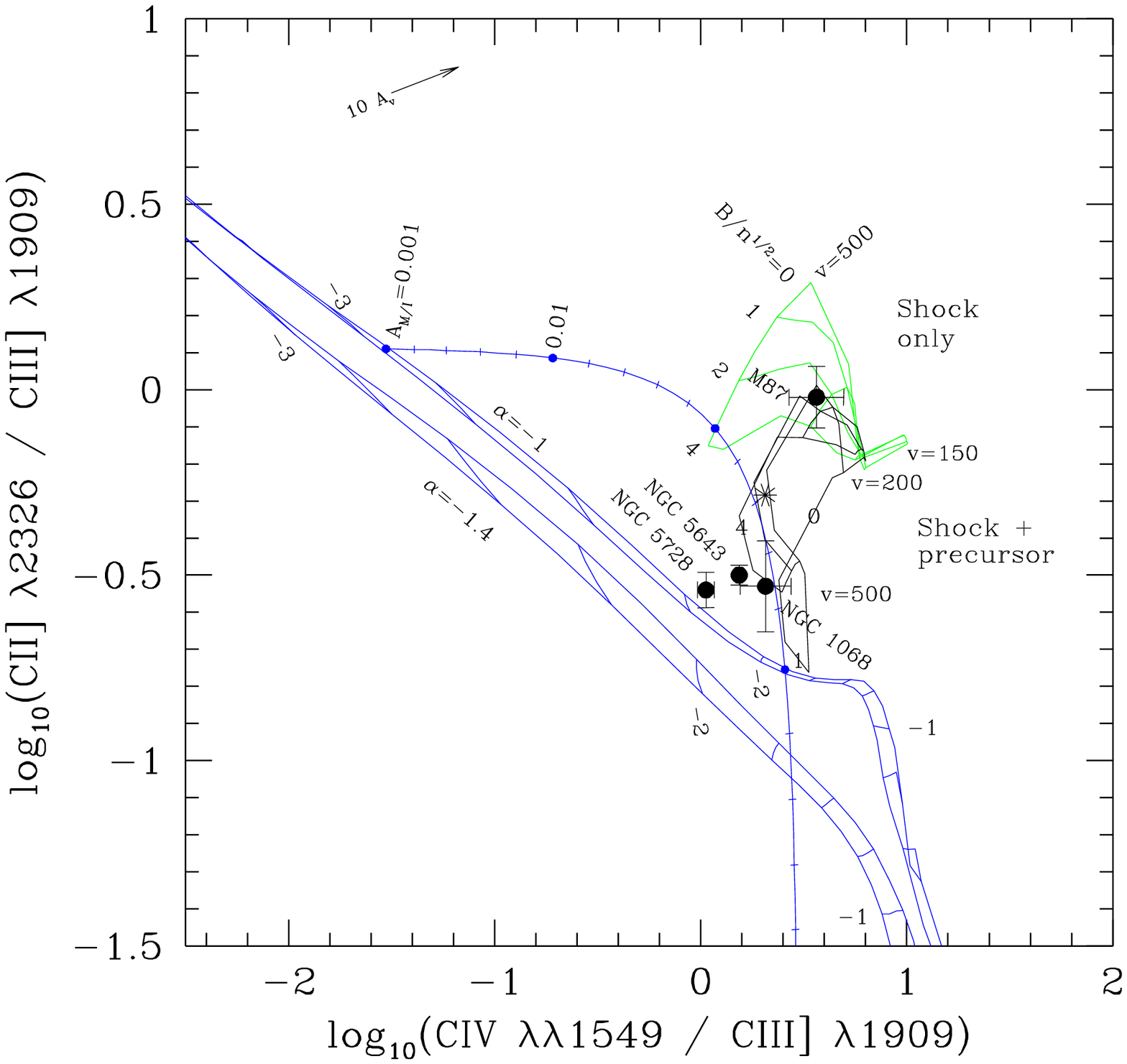}
\caption{UV and UV-optical diagnostic diagrams.}
\end{figure}
 
\clearpage
 
\begin{figure}[tbp]
\plottwo{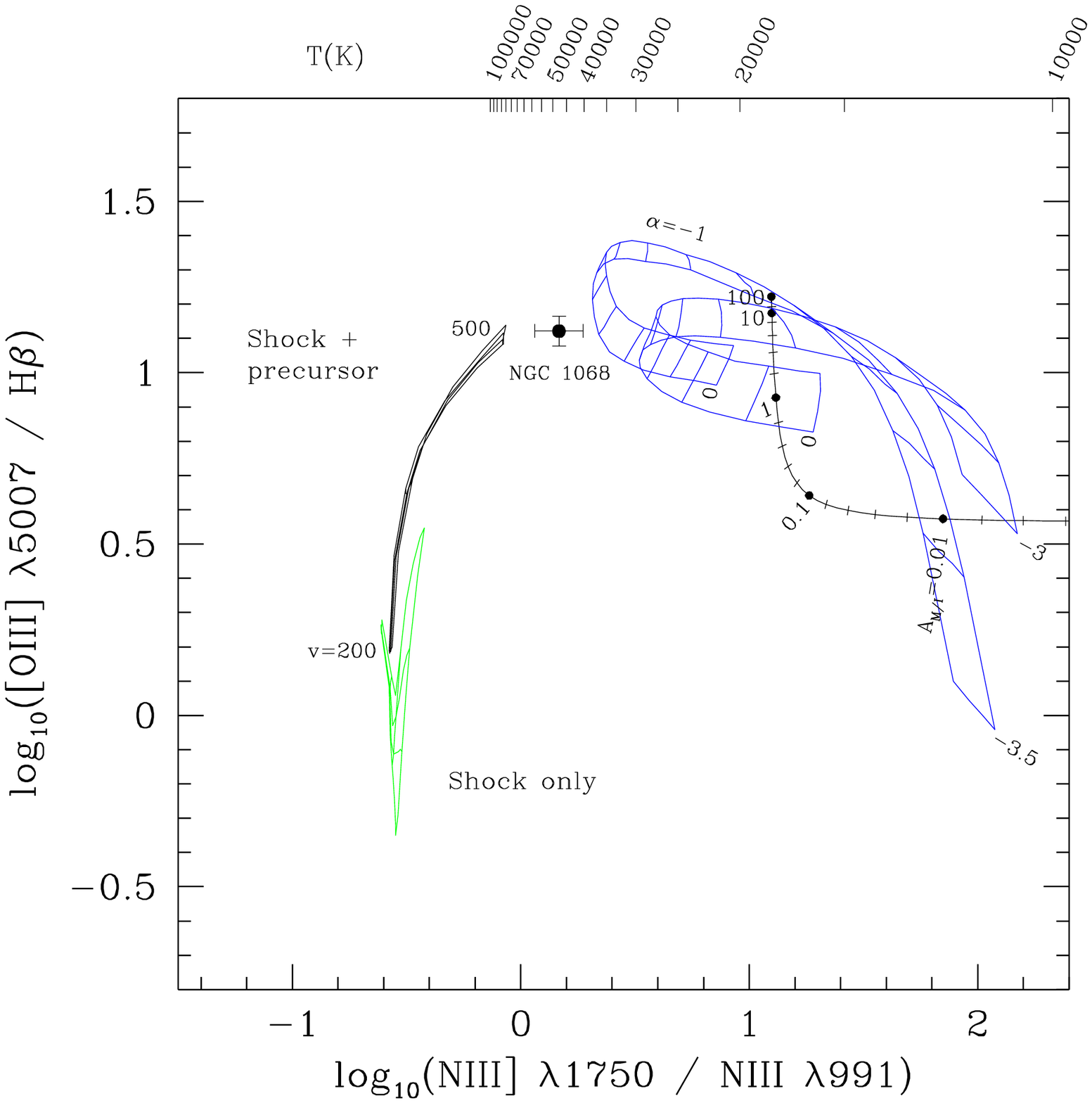}{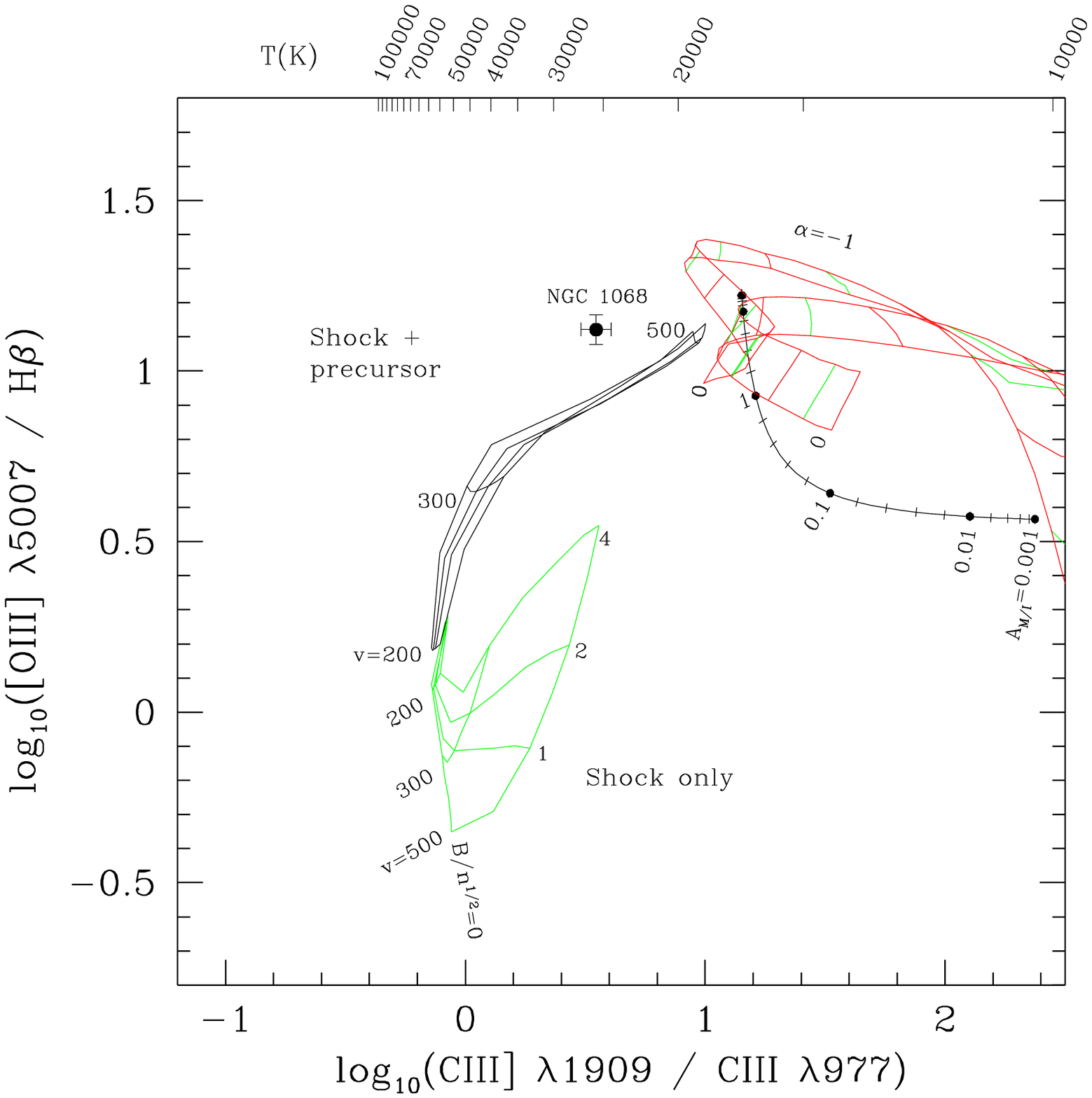}
\caption{Temperature sensitive diagnostic diagrams.}
\end{figure}
  
\clearpage
 
\begin{figure}[tbp]
\plottwo{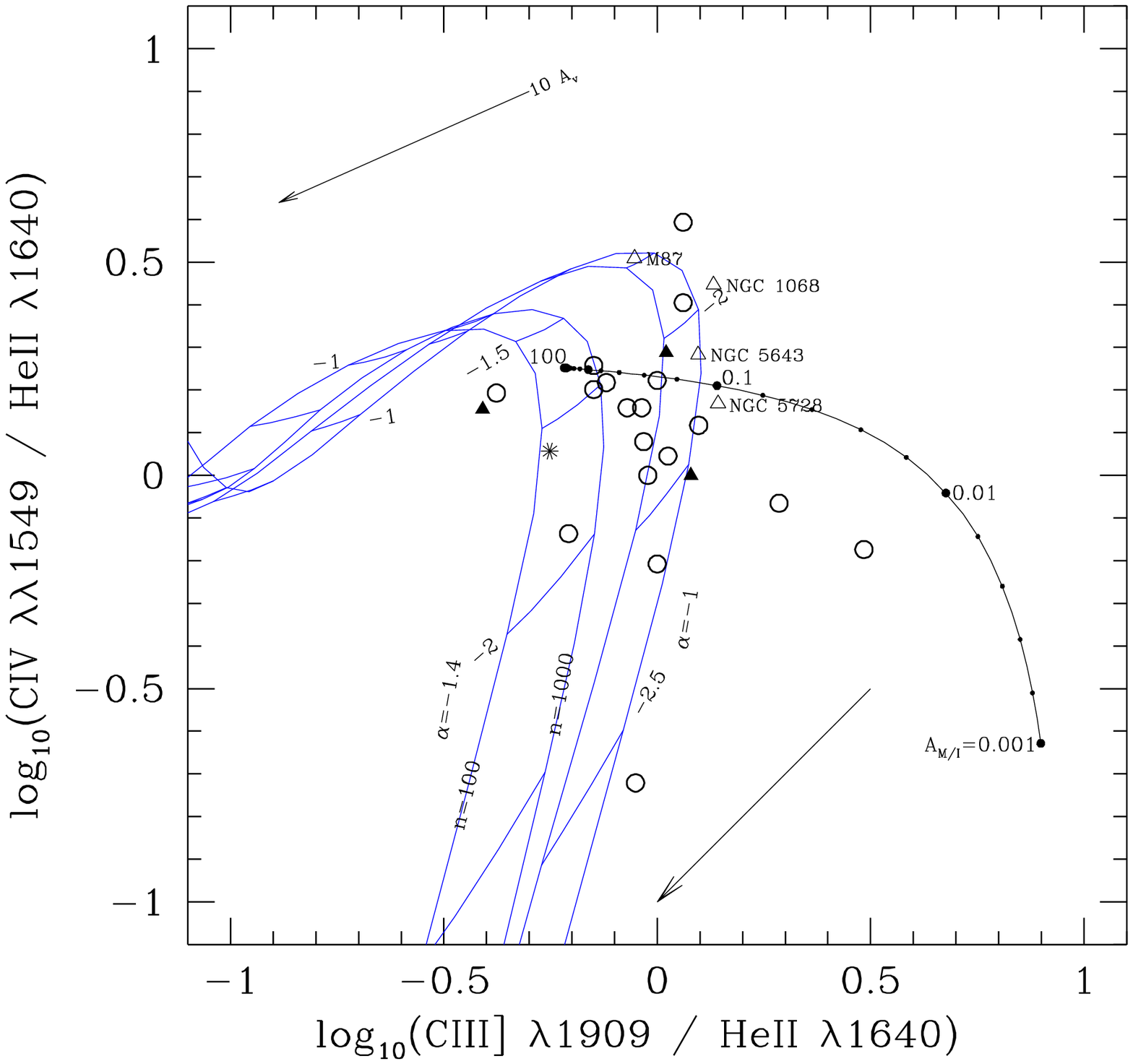}{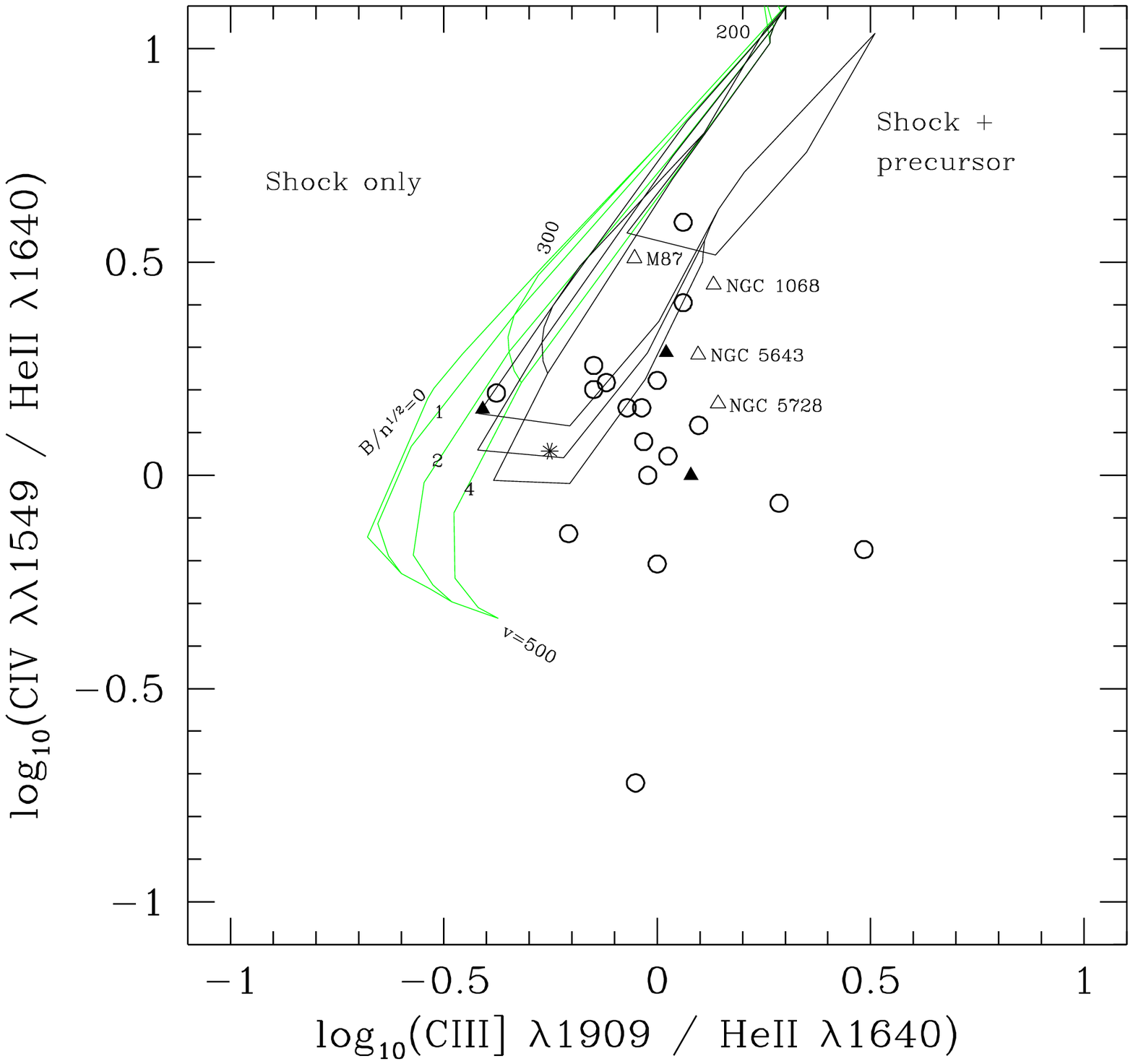} \newline
\plottwo{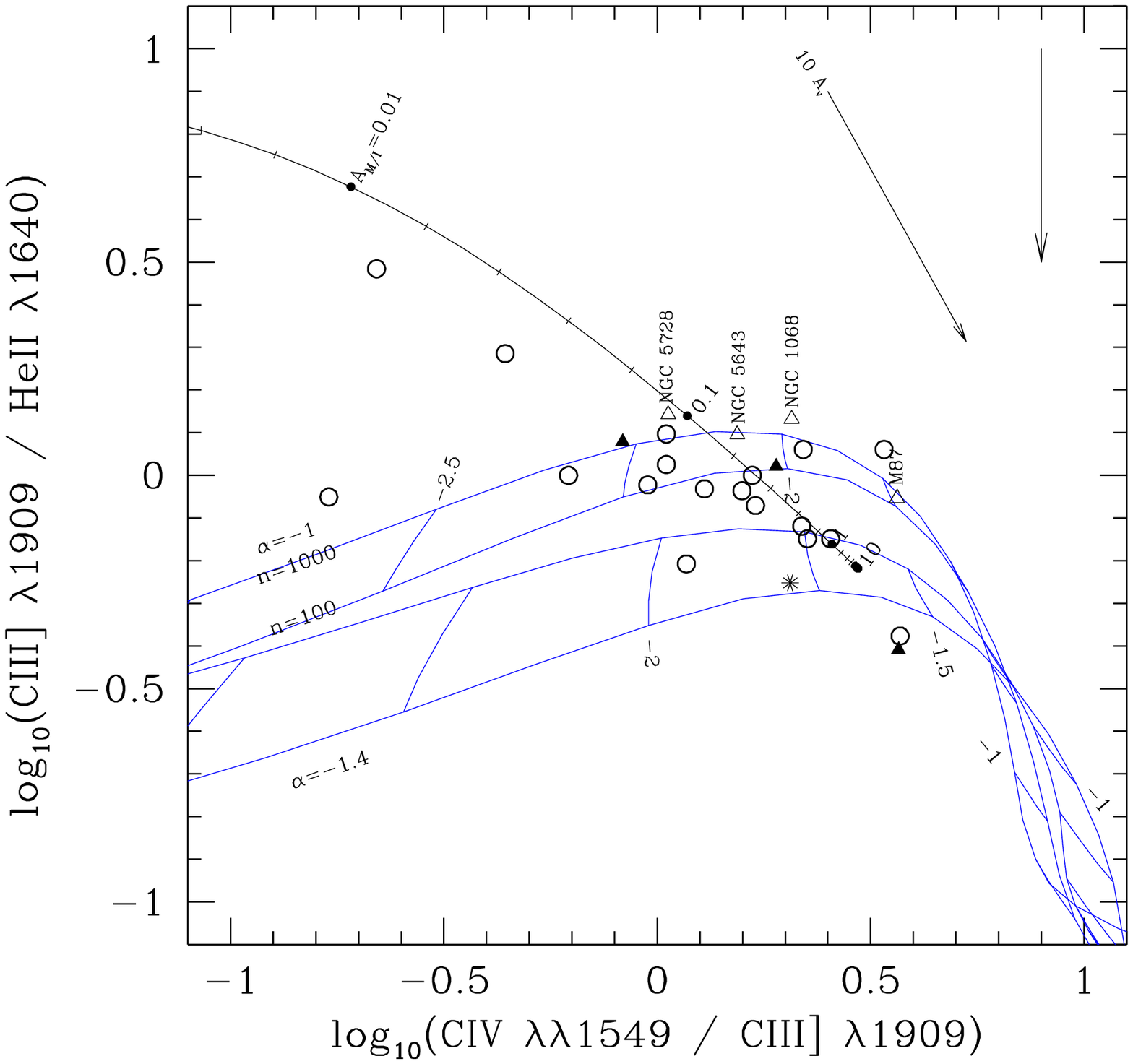}{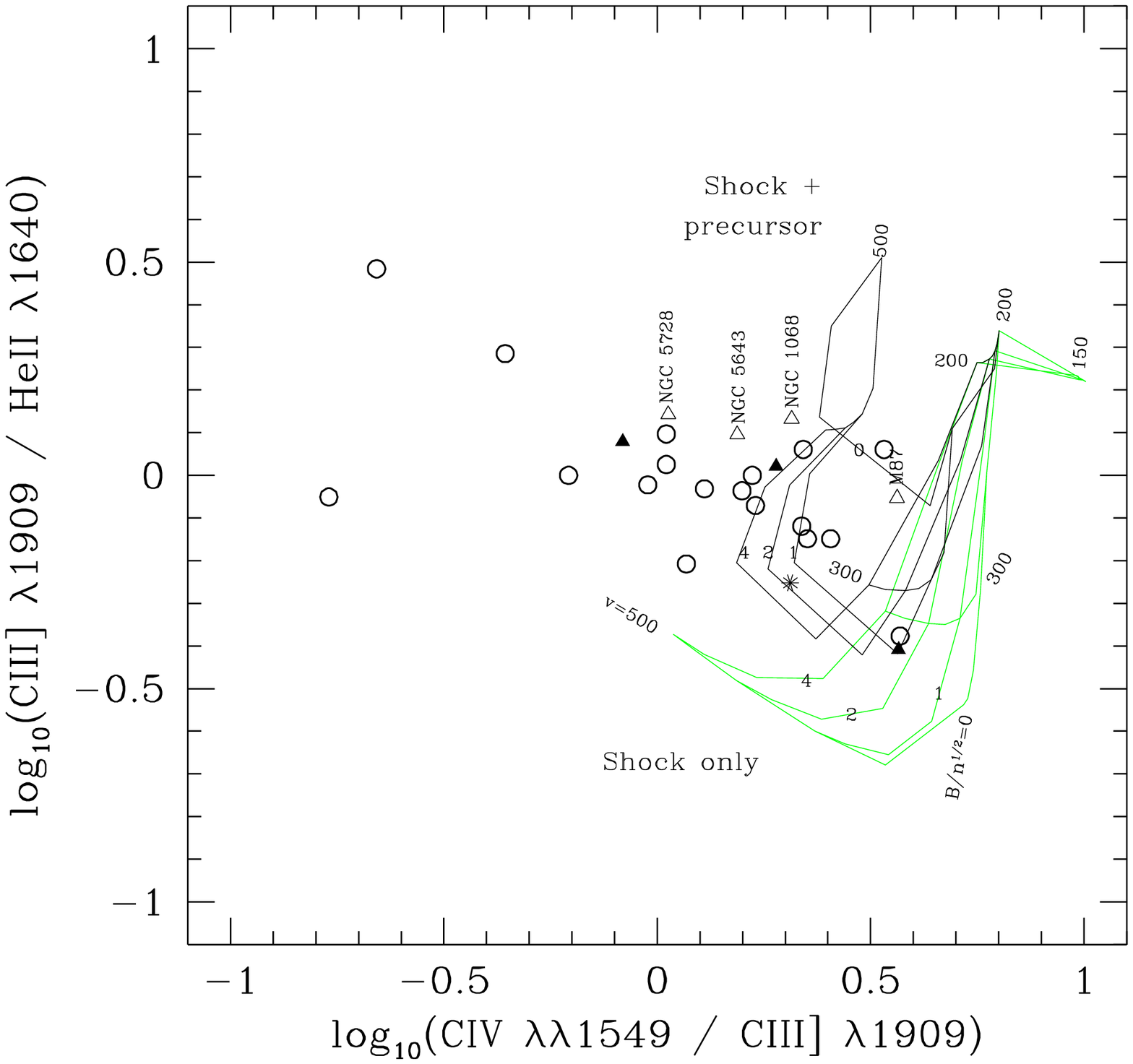} \newline
\plottwo{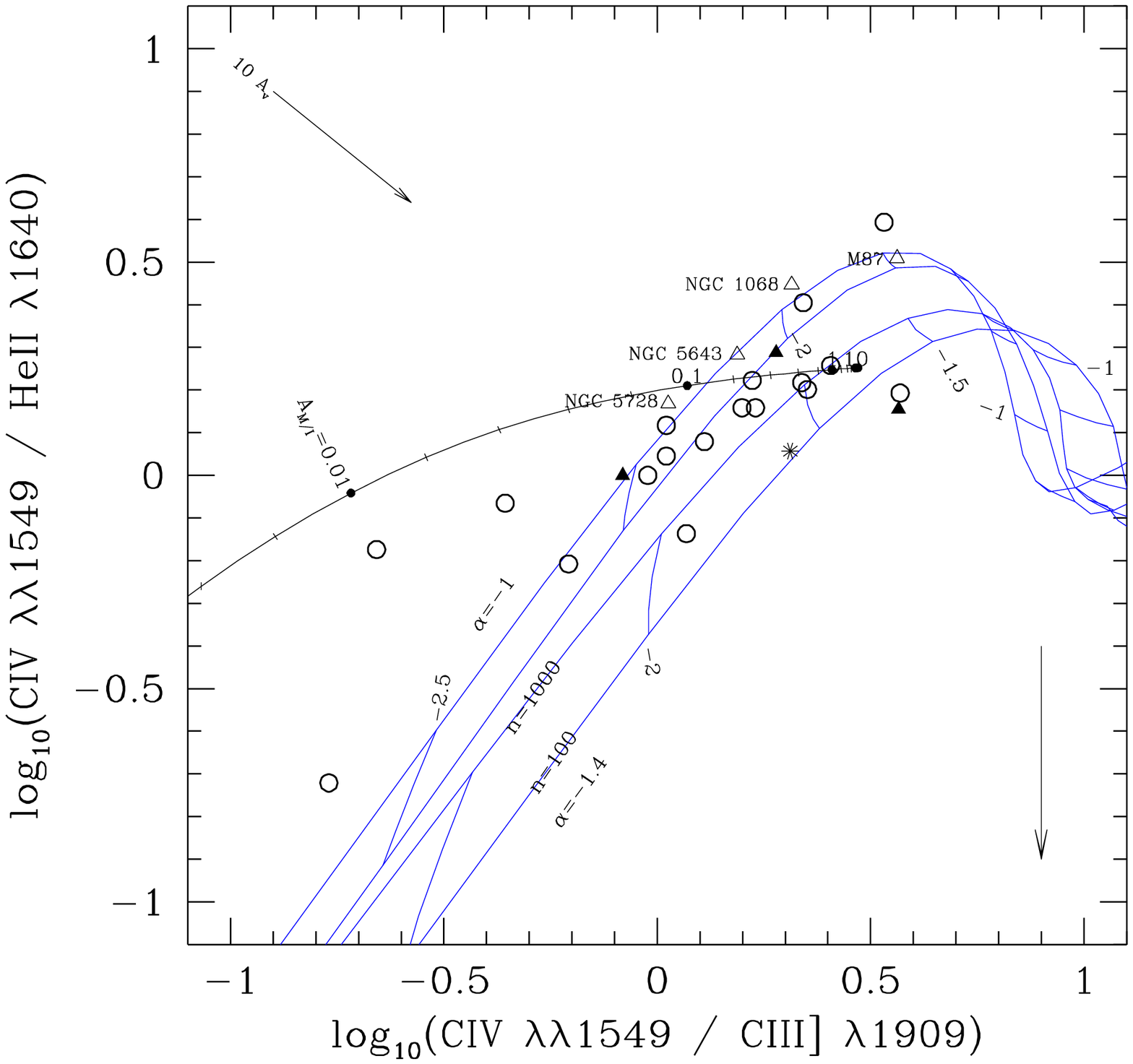}{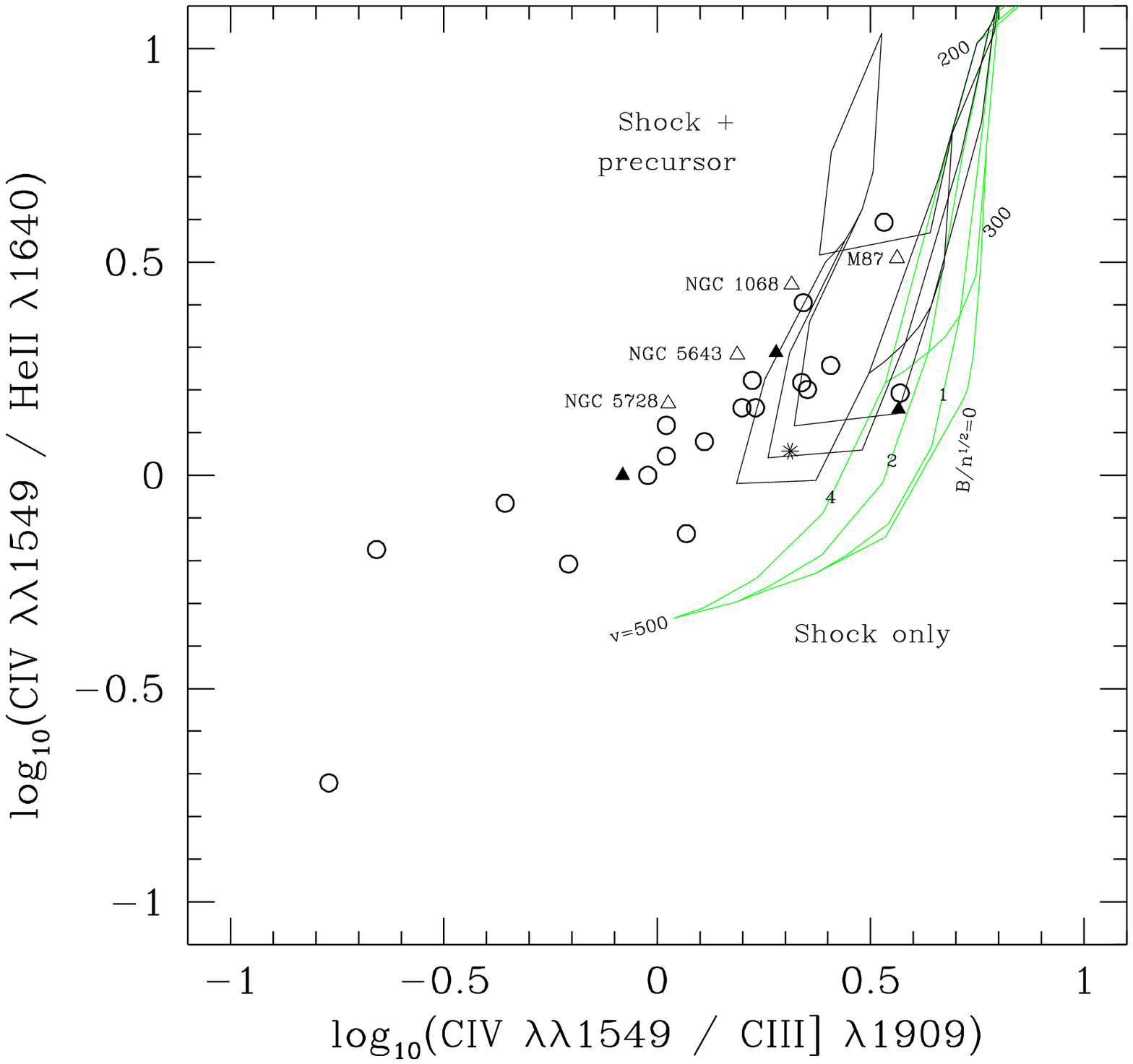}
\caption{UV 
          diagnostic diagrams from Villar-Martin et al. (1997). The open 
            circles are van Ojik's (1995) HZRG and the solid triangles are 
            other published HZRG line fluxes. The star is the average HZRG 
            of McCarthy (1993). }
\end{figure}
 
\end{document}